\documentclass[a4paper,12pt]{article}

\usepackage[english]{babel}
\usepackage{amssymb}
\usepackage{amsbsy}
\usepackage{amstext}
\usepackage{latexsym}
\usepackage{array}
\usepackage{amsmath}
\usepackage{pst-node}

\newcommand {\be} {\begin{eqnarray}}
\newcommand {\ee} {\end{eqnarray}}
\newcommand {\bdm} {\begin{displaymath}}
\newcommand {\edm} {\end{displaymath}}   
\newcommand {\ba} {\begin{array}}
\newcommand {\ea} {\end{array}}
\newcommand {\bse} {\begin{subequations}}
\newcommand {\ese} {\end{subequations}}
\newcommand {\dpa} {\partial}
\newcommand {\pa} [1] {\left(#1\right)}
\newcommand {\paq}[1] {\left[#1\right]}
    
\newcommand {\sca}[1] {\langle #1 \rangle}
\newcommand {\lra} {\leftrightarrow}
\newcommand {\llra} {\longrightarrow}
\newcommand {\lag}{\mathcal{L}}

\newcommand {\Ga}{\Gamma}
\newcommand {\de}{\delta}
\newcommand {\vf}{\varphi}

\newcommand {\Om}{\Omega}
\newcommand {\om}{\omega}
\newcommand {\rn}{\mathbb{R}}
\newcommand {\rr}{\mathcal{R}}
\newcommand {\zn}{\mathbb{Z}}
\newcommand {\na}{\nabla}
\newcommand {\id}{1\!\!1}
\newcommand{\np}[3]{{#1}_{#2}\ldots {#1}_{#3}}
\newcommand{\nn}{\mathcal {\bf N}}

\newcommand{\ep}{\epsilon}
\newcommand{\al}{{\alpha'}}
\newcommand{\si}{\sigma}
\newcommand{\tl}[1]{\tilde {#1}}
\newcommand{\tr}{\mbox{Tr}}

\title{
\begin{flushright}
\normalsize{hep-th/0201166\\
CERN-TH/2002-004\\
HIP-2002-01/TH}
\end{flushright}
\vspace{1cm}
\Large\textbf{Higgs-graviscalar mixing in type I string theory}
\author{I. Antoniadis$^{1}$\footnote{On leave from Ecole Polytechnique CPHT,
UMR du CNRS 7644, F-91128 Palaiseau},\ \ R. Sturani$^{2}$\\ \\
\normalsize $^{1}$ CERN Theory Division, CH-1211 Geneva 23, Switzerland\\
\normalsize $^{2}$ Physics Department, P.O. Box 9, FIN-00014 University of 
Helsinki, Finland}}
\date{}

\begin{document}
\maketitle
\thispagestyle{empty}

\begin{abstract}
We investigate the possibility of mixing between open and closed string
excitations in D-brane models with the fundamental string scale at the TeV.
The open string modes describe the Standard Model Higgs, while closed strings
describe graviscalars living in the bulk. This provides a string setup for
computing the Higgs-graviscalar mixing, that leads to a phenomenologically
interesting invisible width of the Higgs in low scale quantum gravity models,
as suggested previously by Giudice, Rattazzi and Wells.
\end{abstract}
\vspace{1.5cm} 
\hspace{1cm}Keywords: Large Extra Dimensions, Higgs Boson

\newpage

\section{Introduction}

An interesting possibility to address the gauge hierarchy problem and guarantee
its stability is when the string scale lies in the TeV region
\cite{Antoniadis:1998ig, Lykken:1996fj}. In this work, we consider the
scenario of large extra dimensions \cite{Antoniadis:1990ew, 
Arkani-Hamed:1998rs}
in the framework of perturbative type I string theory with the Standard Model 
localized on a
collection of D-branes, in the bulk of $\delta$ extra large compact dimensions
of submillimeter size $R$. Standard Model degrees of freedom are described by
open strings ending on the D-branes, while gravity corresponds to 
closed strings
that propagate also in the bulk.

In this framework we will study the mixing between brane fluctuations, or
\emph{branons} for short, and closed string modes, such as the graviton,
graviphotons and the dilaton or other graviscalars. Since branons are
generically gauge non singlets, such a mixing can arise from 
trilinear couplings
of the form $\si^2 h$, involving two open and one closed string modes that we
denote $\si$ and $h$, respectively. Upon identifying $\si$ with the Standard
Model Higgs scalar, the above coupling induces a Higgs-graviscalar mixing
proportional to the Higgs vacuum expectation value (VEV). It has been suggested
that this mixing leads to an invisible width of the Higgs that may be 
observable
experimentally \cite{Giudice:2000av}. Indeed, since the Higgs is much heavier
than the spacing of the bulk Kaluza-Klein (KK) modes, it would feel a coupling
to a quasi-continuous tower of states, leading to a disappearance amplitude
rather than to oscillations.

In the context of the effective field theory, the required trilinear coupling
$\si^2 h$ was postulated to emerge from an $\mathcal{R}\si^2$ term, where 
$\mathcal{R}$ is the
curvature scalar formed by the \emph{pull-back} metric on the D-brane world
volume. Its coefficient $\xi$ cannot be fixed by the effective field theory and
should be of order unity in order to obtain a visible effect. However, in the
conformal case, one obtains a small value, $\xi=1/6$, dictated by the
conformally coupled scalar in four dimensions.

In this work, we study the branon-bulk mixing in type I string theory and we
compute in particular the trilinear coupling involving two open and one closed
string states. Our results are obtained in supersymmetric theories but remain
valid in non supersymmetric D-brane models, where supersymmetry is 
mainly broken
only on the world-volume of some D-branes, located for instance on top of
anti-orientifold planes \cite{bsb}. More precisely, there are 
three possibilities for the Higgs field that we analyse separately.

In the first case, the Higgs scalar is identified with an excitation of an open
string having both ends on the same collection of parallel D-branes
(\emph{Dirichlet-Dirichlet} or DD open strings in the transverse 
directions). To
lowest order, the effective action can then be obtained by an appropriate
truncation of an $N=4$ supersymmetric theory. In the abelian case, it is given
by the Born-Infeld action, depending on the pull-back of bulk fields on the
D-brane world volume. Expanding in normal coordinates one finds that 
although no $\mathcal{R}\si^2$ term is strictly speaking generated, there is a
quadratic coupling of
branons to the internal components of the Riemann tensor, 
$\rr^\mu_{n\mu m}\si^n\si^m$,
which induces a Higgs-graviscalar mixing. Moreover, there is an additional
coupling of branons with the longitudinal component of a graviphoton in the 
bulk of the form $\si^n\si^m\partial_n\partial_\mu h^\mu_m$. The total effect 
in the invisible width can be summarized in terms of an effective parameter 
$\xi$ which
is of order unity in the case of $\delta=2$ large transverse extra 
dimensions of
(sub)millimeter size. The compatibility of this coupling with the conformal
symmetry of D3-branes can be explained by analyzing the explicit form of the
corresponding conformal transformations.

The second possibility is when the Higgs corresponds to an open string with one
end on the Standard Model branes and the other end on another D-brane extended
in the bulk (\emph{Neumann-Dirichlet} or ND strings). In this case, the branon
interactions do not emerge from a Born-Infeld action but can be extracted
directly by evaluating the corresponding string amplitudes involving twist
fields. An explicit computation of the 3-point function shows that the branon
coupling to the Riemann tensor now vanishes but it remains the mixing with the
graviphoton. As a result, the invisible width is much smaller than in the
previous case.

In the third case, the Higgs lives on a brane intersection, corresponding to an
open string stretched between two orthogonal D-branes transverse to the large
dimensions (ND string in non bulk directions). The Higgs-graviscalar mixing in
this case vanishes.

The paper is organized as follows. In Section~\ref{bis} we consider the first
case where the Higgs is a DD state living on the brane and we derive the
coupling between branons and closed string modes by expanding the Born-Infeld
and Chern-Simons action \cite{Garousi:1998fg}. In Section~\ref{nab} we discuss
the generalization to the non abelian case. In Section~\ref{con} we comment on
the compatibility of the result obtained in the previous sections with the
conformal symmetry of the D3-brane effective action in the $\al\to 0$ limit. In
Section~\ref{mix} we compute the disappearance amplitude for the Higgs. In
Section~\ref{nd} we extend our analysis to the cases where the Higgs emerges as
an excitation of a ND open string, stretched between two orthogonal branes.

\section{Branons' effective action} \label{bis}

In the following we use capital Latin letters for 10-dimensional indices, 
lower case characters ($\mu \ldots\omega$) for indices tangent to D-branes, 
$i\ldots n$ for directions orthogonal to the D-branes, the first part of
Greek alphabet ($\alpha\ldots\delta$) for spinor indices, and $a,b$ will be
used for gauge indices. The metric signature is $(-,+,\ldots,+)$.

We start by considering the effective field theory on a single D$p$-brane,
i.e. with $U(1)$ gauge group, which is given by the sum of Born-Infeld and 
Chern-Simons actions:
\be \label{bi}
S_{BI}&=&-T_p\int d^{p+1}x\, e^{-\tl\Phi}\sqrt{-\det\pa{\tl G_{\mu\nu}+
\tl B_{\mu\nu}+ 2\pi\al F_{\mu\nu}}}\\
\label{cs}
S_{CS}&=&\mu_p\int d^{p+1}x\, e^{\tl B+2\pi\al F}\wedge\sum_{p}\tl C^{(p+1)}\ ,
\ee
where $F_{\mu\nu}$ is the field strength of the abelian
world-volume gauge field, and $T_p$, $\mu_p$ are the tension and Ramond-Ramond
(R-R) charge of the D$p$-brane. The closed string fields are the 
\emph{pull-back} of the bulk fields to the D-brane world volume:
\bse \label{pulbec}
\be
&&\tl G_{\mu\nu}=G_{\mu\nu}+G_{m\mu}\dpa_\nu \si^m+G_{m\nu}\dpa_\mu\si^m+
G_{mn}\dpa_\mu\si^m\dpa_\nu\si^n\\
&&\tl B_{\mu\nu}=B_{\mu\nu}+B_{m\mu}\dpa_\nu \si^m-B_{m\nu}\dpa_\mu\si^m+
B_{mn}\dpa_\mu\si^m\dpa_\nu\si^n \\
&&\tl \Phi\ =\Phi\\
&&\tl C^{(p+1)}_{\np{\mu}{0}{p}}=C^{(p+1)}_{\np{\mu}{0}{p}}+
\dpa_{\mu_0}\si^mC^{(p+1)}_{m\np{\mu}{1}{p}}+\dpa_{\mu_0}\si^m\dpa_{\mu_1}
\si^nC^{(p+1)}_{mn\np{\mu}{2}{p}}\ ,
\ee
\ese
where $G$, $B$, $\Phi$ and $C$ are the metric, two index antisymmetric tensor, 
dilaton and the R-R $(p+1)$-form potential, respectively.
Here, we define the transverse coordinates of the brane as our $\si$ fields 
and an implicit antisymmetrization over indices $\mu_0,\mu_1,\ldots,\mu_p$ in 
(\ref{pulbec}d) is understood.

The tension and R-R charge of the brane can be computed either by T-dualizing
the tree level one-point closed string amplitude \cite{Ohta:1987nq} or by  
one loop vacuum computation \cite{Polchinski} and they are given in terms of
the Yang-Mills coupling $g_{YM}^{(p)}$ on the brane by
\be
\mu_p=T_p=\pa{2\pi\al g_{YM}^{(p)}}^{-2}\ .
\ee
We now recast the Born-Infeld action (\ref{bi}) into the Einstein 
frame, where the Ricci scalar in the bulk action is canonically normalized. 
This is obtained by rescaling to the Einstein metric $g_{MN}$ given by
\be
g_{MN}=e^{-\Phi/2}G_{MN}\, ,
\ee
in terms of which the corresponding action $S_{BI}^{(E)}$ is
\be \label{bie}
S_{BI}^{(E)}=-T_p\int d^{p+1}x\ e^{\frac{p-3}{4}\tl\Phi}\sqrt{-\det
\pa{\tl g_{\mu\nu}+e^{-\tl\Phi/2}\tl B_{\mu\nu}+ 2\pi\al e^{-\tl\Phi/2} 
F_{\mu\nu}}}\ .
\ee
No rescaling is needed for the Chern-Simons action as it is metric
independent. 

Expanding $S^{(E)}_{BI}+S_{CS}$ around a flat Minkowski space,
\bse
\be
g_{MN}&=&\eta_{MN}+h_{MN}\\
B_{MN}&=&b_{MN}\\
\Phi&=&\phi\ ,
\ee
\ese
one obtains \cite{Garousi:1998fg}
\renewcommand{\arraystretch}{1.8}
\be \label{ac} 
&&\lag^{(1)}=-T_p \paq{\frac{(p-3)}{4}\phi+\frac{1}{2}
h^\mu_\mu}\pm\mu_p C_{\np{\mu}{0}{p}}^{(p+1)}
\frac{\ep^{\np{\mu}{0}{p}}}{(p+1)!}\ ,\\
&&\ba{ll} \label{1op1cl}
\lag^{(2)}=-T_p\displaystyle \paq{\dpa_\mu\si^m h^\mu_{\ m}+
\si^m\frac{\dpa_mh^\mu_\mu}{2}+\frac{p-3}{4}\si^m\dpa_m\phi
+\pi\al b_{\mu\nu}F^{\mu\nu}} \\ 
\qquad\quad \pm \mu_p\displaystyle\left[\pa{\si^m \dpa_m 
C^{(p+1)}_{\np{\mu}{0}{p}}+(p+1)\si^m 
\dpa_{\mu_0}C^{(p+1)}_{m\np{\mu}{1}{p}}}\frac{\ep^{\np{\mu}{0}{p}}}{(p+1)!}
+\right.\\
\qquad\qquad\qquad\displaystyle \pi\al F\wedge C^{(p-1)}\Big]\ ,
\ea\\
&&\lag^{(3,NS^2)}=\left\{\frac{1}{2g^2_{YM}}\Big[
\pa{\dpa_\rho A_\mu\dpa^\rho A_\nu + \dpa_\mu A^\rho\dpa_\nu A_\rho 
-2\dpa_\mu A^\rho\dpa_\rho A_\nu}h^{\mu\nu}-\right.\nonumber\\
&&\qquad\qquad\left.\pa{\dpa_\rho A_\si \dpa^\rho A^\si 
-\dpa_\rho A^\si \dpa_\si A^\rho}\frac{h^\mu_\mu}2-
\pa{\dpa_\rho A_\si \dpa^\rho A^\si-\dpa_\rho A^\si \dpa_\si A^\rho}
\frac{p-7}{4}\phi\right]\nonumber\\
&&\qquad\qquad +\frac{T_p}{2}\left[\pa{\dpa_\mu\si_m\dpa_\nu\si^m
-\frac{1}{2}\eta_{\mu\nu}\dpa_\rho\si_m\dpa^\rho\si^m}h^{\mu\nu}-
h_{mn}\dpa_\rho \si^m\dpa^\rho \si^n- \right. \label{2ph1g}\\ 
&&\qquad\qquad\dpa^\rho \si^m\dpa_\rho\si_m
\ \pa{\frac{p-3}{4}}\phi-\si^m\si^n\dpa_m\dpa_n\pa{\frac{p-3}{4}\phi+
\frac{h^\mu_\mu}{2}}-\nonumber \\
&&\qquad\qquad 2\pa{\dpa_\mu\si^n}\si^m\dpa_m h_n^\mu-
2\pi\al \pa{2 b_{\mu m}F^{\mu\nu}\dpa_\nu\si^m+
\si^mF_{\mu\nu}\dpa_m b^{\mu\nu}}\Big]\Big\}\ ,\nonumber\\
&&\ba{ll} \label{2ph1c}
\lag^{(3,R^2)}_{br}=\pm\mu_p\displaystyle
\left\{\frac{1}{2}\si^m\si^n\dpa_m\dpa_nC^{(p+1)}_{\np{\mu}{0}{p}}
+(p+1)\si^m\dpa_{\mu_0}\si^n\dpa_mC^{(p+1)}_{n\np{\mu}{1}{p}}+\right.\\
\qquad\quad\displaystyle \frac{(p+1)p}{2}\dpa_{\mu_0}\si^m\dpa_{\mu_1}\si^n 
C^{(p+1)}_{mn\np{\mu}{2}{p}}+
\frac{(p+1)p}{2}(2\pi\al)F_{\mu_0\mu_1}\si^m\dpa_mC^{(p-1)}_{\np{\mu}{2}{p}}+\\
\displaystyle\quad\qquad
\frac{(p+1)p(p-1)}{2}(2\pi\al)F_{\mu_0\mu_1}\dpa_{\mu_2}\si^m
C^{(p-1)}_{m\np{\mu}{3}{p}}+\\ \displaystyle\quad\qquad
\left.\frac{(p+1)p(p-1)(p-2)}{8}(2\pi\al)^2F_{\mu_0\mu_1}F_{\mu_2\mu_3}
C^{(p-3)}_{\np{\mu}{4}{p}}\right\}\frac{\ep^{\np{\mu}{0}{p}}}{(p+1)!}\, ,
\ea
\ee
\renewcommand{\arraystretch}{1}
\hskip -.2cm where the $\pm$ signs correspond to the two choices of the
D-branes R-R charge (branes or anti-branes) and $\ep^{\np{\mu}{0}{p}}$ is
the usual  antisimmetric tensor density.
 
The non kinetic terms in the above expressions (with no spacetime derivative 
on $\si$) are obtained by retaining the terms up to quadratic level of the 
Taylor expansion  
\be \label{taysi}
\sum_{k=1}^\infty\frac{\pa{\si^m\dpa_{y^m}}^k}{k!}
\pa{e^{\frac{p-3}{4}\Phi}\sqrt g \mp C^{(p+1)}}
(y^n)\Big|_{y_m=0}\ .
\ee
This shows that the branons experience a non derivative interaction in a 
nontrivial background, which can be interpreted as a potential $V_{br}$ 
for the position of the brane 
\be \label{alpo}
V_{br}\equiv T_p\pa{e^{\frac{p-3}{4}\Phi}\sqrt g\mp C^{(p+1)}}\ .
\ee
We expect that in a supersymmetric background the Neveu-Schwarz Neveu-Schwarz
(NS-NS) and the Ramond Ramond (R-R) fields give mutually compensating 
contribution to the potential term: we shall check this fact in 
Section.~\ref{con}, using the supergravity description of branes.

Let us consider now the trilinear Lagrangian (\ref{2ph1g}). It corresponds to 
the closed string linearization of the following non linear 
Lagrangian, quadratic in NS open string modes:
\renewcommand{\arraystretch}{1.8}
\be \label{azns}
\ba{c} 
\lag^{NS}=-\displaystyle  e^{\frac{p-3}{4}\Phi}\sqrt{g}
\left[\frac{1}{4g_{YM}^2}F_{\mu\nu}F_{\rho\sigma}g^{\mu\rho}g^{\nu\sigma} 
+\frac{T_p}{2} \left(\na_\mu\si^m\na_\nu\si^n g^{\mu\nu}g_{mn}-\right.\right.\\
\qquad\qquad\left.\si^m\si^n\rr^\mu_{\ m\mu n}+2\pi\al\pa{2b^\mu_{\ m} 
F_{\mu\nu}\dpa^\nu\si^m+F_{\mu\nu}\si^m\dpa_mb^{\mu\nu}}\right)\Big]\, ,
\ea
\ee
\renewcommand{\arraystretch}{1}
\hskip -.2cm where we introduced the (gravitational) covariant derivative over
$\si$ fields
\be
\na_\mu \si^m= \dpa_\mu \si^m + \Ga^m_{n\mu}\si^n\ .
\ee 
The gravitational connection is given by
\be
\Ga^m_{n\mu}=\frac{1}{2}g^{mM}\pa{g_{nM,\mu}+g_{M\mu,n}-g_{n\mu,M}}\ ,
\ee
where column denotes differentiation as usual.

We thus found, besides the expected Yang-Mills kinetic terms, a potential
of interaction between branons and the bulk closed string
states. Note that the potential term in (\ref{azns}) vanishes in a trivial 
background; it generates interactions of $\si^m$ with higher KK modes of the 
bulk fields. The above results can be also obtained by a direct computation of 
corresponding on shell string amplitudes.

\section{Non-abelian generalization} \label{nab}

In the non-abelian case, we cannot rely on the Born-Infeld action to obtain
the  effective field theory. Instead, one can compute the relevant 3-point
amplitude  involving two branons and one closed string state.
We concentrate below on the NS-NS sector.
The amplitudes are given in terms of one kinematical 
invariant variable $t$, which is given in terms of the momenta of
the open string excitations, $k_2$, $k_3$, and the momentum $k_1$ of the 
closed string state by
\be \label{kine}
(k_2+k_3)^2=2k_2\cdot k_3=-t \nonumber\\ \label{kine21}
k^\mu_1\cdot k_{1\mu}=-k^m_1\cdot k_{1m}=-t\\
(k_2+k_1)^2=(k_3+k_1)^2=t\, , \nonumber
\ee
where the last product is understood to be over the full ten dimensional space,
such that $k_{2,3}$ have non-vanishing components along the brane directions
only, and $\sqrt{-t}$ is the KK mass of the closed string particle. The
relevant  amplitudes in our analysis are\footnote{The dilaton 
is defined by a polarization tensor $\zeta_{MN}$ of the closed string vertex 
operator of the form $\zeta_{MN}=\phi (\eta_{MN}-l_Mk_N-k_Ml_N)/\sqrt 8$, where
$k$ is its momentum and $l$ a vector satisfying $l^2=0$, $kl=1$.} 
\cite{Hashimoto:1996bf}
\renewcommand{\arraystretch}{1.6}
\bse \label{a2op1cl}
\be &&
\ba{ll}
A(h,\si,\si)&=ig_c\frac{2^{-\al t}\sqrt\pi\Ga(1/2-\al t/2)}{\Ga(1-\al t/2)}
\tr(t^at^b)\left[2ik_{2\mu}\si^{am}ik_{3\nu}\si^b_mh^{\mu\nu}\right.\\
&-2ik_{2\mu}\si^{am}ik_3^\mu\si^{bn}h_{mn}-
4ik_{2\mu}\si^{an}\si^{bm}ik_{1m}h^\mu_n-\\
&\left.\si^{am}\si^{bn}ik_{1m}ik_{1n}h^\mu_\mu
-ik_{2\mu}\si^{am} ik_3^\mu\si_m^bh^\nu_\nu(1+\al t)^{-1}\right]+(2\lra 3)
\ea \\
&&\ba{ll}
A(\phi,\si,\si)&=ig_c\frac{2^{-\al t}\sqrt\pi\Ga(1/2-\al t/2)}{\Ga(1-\al t/2)}
\tr(t^at^b)\left[\si^{am}\si^{bn}ik_{1m}ik_{1n}(3-p)+ \right.\\
&\left. ik_2\si^{am}ik_3\si^b_m\pa{\frac{4-p}{1+\al t}-1}
\right] \times \frac{1}{2\sqrt 2}\phi + (2\lra 3)
\ea 
\ee
\ese
\renewcommand{\arraystretch}{1}
\hskip -.2cm which are in agreement with the Lagrangian (\ref{azns}) in the
$U(1)$ case and  justify its straightforward generalization to nontrivial
Chan-Paton factors.  Comparing the two expressions one must use the relations
\cite{Polchinski}
\be \label{go}
g_c&=&\kappa /2\pi 
\ee
and the following rescaling of the fields
\renewcommand{\arraystretch}{1.5}
\bse \label{resca}
\be 
\si^m &\to & \si^m/\pa{2\pi\al g_{YM}}\ ,\\
h_{MN} &\to & h_{MN}/(2\kappa)\ ,\\
\phi &\to & \sqrt 2\phi/(2\kappa)\ ,
\ee
\ese
\renewcommand{\arraystretch}{1}
where $\kappa=\sqrt{8\pi G_N}$ is the 
gravitational coupling.

\section{Conformal invariance} \label{con}

It is known that the gauge field theory on a D$3$-brane is $\nn=4$ 
supersymmetric, which is conformal invariant. It is also known that 
conformally coupled scalar fields in four dimensions exhibit a 
$\xi\mathcal{R}\si^2$ 
interaction with $\xi=1/6$ and $\mathcal{R}$ the four-dimensional Ricci scalar.
One may be worried why the calculations exposed so far do not show the 
expected $\xi\mathcal{R}\si^2$ coupling for $p=3$, which is also a source of 
Higgs-graviscalar mixing. In fact we shall argue below that the conformal
symmetry is realized in a rather different way. Moreover in Section~\ref{mix}
we shall show that the potential interaction we obtained in Eq.~(\ref{azns})
gives rise still to a disappearance amplitude for the Higgs that can be
parametrized in terms of an effective $\xi$ that we shall compute.

Looking back to (\ref{azns}), we can check that conformal invariance of the
gauge fields for $p=3$ (and no other values of $p$) is obtained in the usual 
way, using a conformal transformation on the brane 
\be \label{cobra}
g_{\mu\nu}\to \hat g_{\mu\nu}=\Omega^2 g_{\mu\nu}\ ,
\ee
with the dilaton $\Phi$ and the gauge field $A_\mu$ inert. For the scalar 
branons the situation is different as they also couple to 
graviphotons and graviscalars. Here we show that conformal invariance on a 
flat $3$-brane is achieved if (\ref{cobra}) is supplemented by 
\be \label{cobra2}
g_{mn}\to \hat g_{mn}=\Omega^{-2} g_{mn}\ ,
\ee
with the branons unaltered. Indeed, it is easy to see that (\ref{cobra}) and 
(\ref{cobra2}) applied together correspond to a conformal symmetry of the 
action derived from (\ref{azns}) for $p=3$ in a trivial background. Moreover 
in the presence of a R-R field, applying the transformations
\bse \label{cobra3}
\be
C_{\np{\mu}{0}{p}}\to \hat C_{\np{\mu}{0}{p}}&=&\Om^4C_{\np{\mu}{0}{p}}\\
C_{m\np{\mu}{1}{p}}\to \hat C_{m\np{\mu}{1}{p}}&=&\Om^2C_{m\np{\mu}{1}{p}}\\
C_{mn\np{\mu}{2}{p}}\to \hat C_{mn\np{\mu}{2}{p}}&=&C_{mn\np{\mu}{2}{p}}\, ,
\ee
\ese
with $g_{\mu m}$ inert, one can show that the conformal symmetry is 
exact provided the background is chosen so that the potential 
(\ref{alpo}) vanishes and that 
$\sqrt{g}h^{\mu_0}_m=C_{m\np{\mu}{1}{p}}\ep^{\np{\mu}{0}{p}}/p!$. Here, we
dropped for simplicity the $(p+1)$ superscript on the R-R form $C^{(p+1)}$.

The effect of the $\xi\mathcal{R}\si^2$ term is thus replaced by other 
interaction
which at the quadratic level becomes $\si^m\si^n\rr^\mu_{\ m\mu n}$. 
In relation
to these effects, one may wonder about the argument in \cite{Witten:1998qj}, 
where the $\nn=4$ super Yang-Mills theory was considered on $S^4$ rather then
on $\rn^4$  and it was claimed that the flat direction of $\si$ is lifted by
the curvature coupling $\mathcal{R}\si^2$, thus making the path integral to 
converge,
at least if the metric is close enough to the one of a four sphere, which has
$\mathcal{R}>0$. In our case the $(\si^m\dpa_m)^k(\sqrt{g}+C)$ or its quadratic
expansion may equally well do the job for the case of a four sphere embedded
in a higher dimensional spacetime. It would be interesting to check this
explicitly.

Amusingly enough, there is at least one case in which the potential 
(\ref{alpo}) vanishes in a non-trivial way, thus preserving the conformal
symmetry. It corresponds to the supergravity background induced by some
parallel 
D$p$-branes. The background is given, in the
string frame, for $p<7$, by \cite{Duff:1994an}
\renewcommand{\arraystretch}{1.7}
\be \label{sbrane}
\ba{l}
ds^2=H^{-\frac{1}{2}}\pa{dx^\mu dx_\mu}+ H^{\frac{1}{2}} \pa{dy^mdy_m}\\
e^\Phi=H^{\frac{p-3}{4}}\\
C_{\mu_0\ldots\mu_p}=\ep_{\mu_0\ldots\mu_p}H^{-1}\\ \displaystyle
H=1+\frac{Q}{7-p}\frac{1}{r^{7-p}}
\ea
\ee
\renewcommand{\arraystretch}{1}
\hskip -.2cm where $r\equiv \sqrt {y^my_m}$. The solution (\ref{sbrane})
depends on the  parameter $Q$, with dimensions of $[length]^{7-p}$, defined by
\be \label{Q}
\int_{\perp} d*dC^{(p+1)}=\int_{\dpa\perp}*dC=S_{8-p}Q
\ee
where $S_{8-p}$ is the volume of the $(8-p)$-dimensional sphere of unit 
radius. The classical parameter $Q$ is related to the microscopic string 
parameter $\mu_p$ by\footnote{This is derived using the terms in 
the supergravity action that involve the R-R form 
\bdm
S=-\frac{1}{4\kappa^2}\int d^{10}x (dC)^2 - \mu_p\int_{branes} C\ .
\edm
Thus, the equation of motion is
\bdm
\int_\perp d*dC=2\kappa^2\sum_{branes}\mu_p 
\edm
which, compared to (\ref{Q}), gives (\ref{Qmu}).}
\be \label{Qmu}
Q=N\mu_p 2\kappa^2/S_{8-p}
\ee
where the integer $N$ counts the number of D-branes and the 
classical limit is recovered at $N\to \infty$.
Using that $\mu_p\propto g_o^{-2}$, $\kappa\propto g_c$ and that
$g_o^2\propto g_c\propto e^{\langle\Phi\rangle}\equiv g_s$
we have also
\be \label{Qgs}
Q\sim Ng_s(2\pi\al)^{\frac{7-p}{2}}\ .
\ee

In this supersymmetric background the potential felt by a stuck of $N'$ test 
D$p$-branes with tension and charge $T^{test}_p$ (assuming $N\gg N'$ so that 
the test branes alter negligibly the background they are plunged into) is
\be \label{noforce}
V_{br}=T_p H^{-1}-\mu_p H^{-1}=0
\ee
where we used the BPS relation $T_p=\mu_p$.
We also note that in this case the conformal transformations 
(\ref{cobra}), (\ref{cobra2}) and (\ref{cobra3}) can be described at once by 
defining the conformal transformation of the function $H$ 
\be \label{concur}
H(r)\to \hat H(r)=\Omega^{-4}H(r)\ .
\ee
If the background is non supersymmetric and cancellation (\ref{noforce}) does 
not hold (as in the case of a test antibrane) conformal invariance is 
broken.

Finally, we check the limit of validity of the supergravity solution
(\ref{sbrane}). We should have $\al R\ll 1$ for the curvature scale $R$, 
and the weak coupling condition $e^{\Phi}\ll 1$. In fact, on the background
(\ref{sbrane}) we have
\be
R&\sim& \frac{{H'}^2}{H^{5/2}}\left\{
\ba{lc} 
\stackrel{r\to 0}{\llra} & Q^{-1/2}r^{\frac{3-p}2}\\
\stackrel{r\to\infty}{\llra} & \pa{Q/r^{8-p}}^2
\ea\right. \nonumber \\
e^\Phi &=& H^{\frac{p-3}{4}}\left\{
\ba{lc} 
\stackrel{r\to 0}{\llra} & \pa{\frac{Q}{7-p}}^{\frac{p-3}{4}}
r^{\frac{(p-7)(p-3)}{4}}\\
\stackrel{r\to\infty}{\llra} & 1+\frac{p-3}{4(7-p)}\frac{Q}{r^{7-p}}
\ea\right. \nonumber
\ee
Thus, in the $r\to\infty$ limit the curvature vanishes and the coupling is 
bounded for every $p$, whereas in the $r\to 0$ limit both curvature and 
coupling blow up for $p>3$. However, in the $p\leq 3$ case, the curvature is
bounded by $R_{max}$
\be
R_{max}\propto Q^{-\frac{2}{7-p}}
\ee
and thus, for $p\leq 3$, $\al$ corrections can be taken under control for any 
value of $r$ by sending $Q\to\infty$ in a way that $\al Q^{-2/(7-p)}\to 0$
(for $p=3$, the $AdS_5\times S_5$ geometry is obtained in this way).
If we plonge a brane into a nontrivial general background, relation 
(\ref{noforce}) generally won't hold and a potential for the position of the 
brane will be generated. 

Hence, everything appears to be consistent even in the absence of a 
$\xi\mathcal{R}\si^2$ term.

\section{Higgs-graviscalar mixing} \label{mix}

We shall now show how in our scenario a mixing may take place between branons
and a graviscalar. The mixing is triggered by the trilinear coupling 
$\si^2 h$ in (\ref{2ph1g}) if $\si$ acquires an expectation value
\cite{Giudice:2000av}. Before we analyse the mixing, we discuss first the
abelian case of a single brane, where the graviphoton absorbs the branon and
acquires a (localized) mass. For this purpose we need the expansion of
the Born-Infeld action (\ref{bie}) at the quadratic level of the
NS-NS closed string modes:
\renewcommand{\arraystretch}{1.5}
\be \label{2nsns}
\ba{ll} \displaystyle
\lag_{2NS^2}=-T_p &\paq{\frac{1}{8}\pa{h^\mu_\mu}^2
-\frac{1}{4}h_{\mu\nu}h^{\mu\nu}+\frac{1}{2}\pa{\frac{p-3}{4}}^2\phi^2+
\frac{p-3}{8}\phi h^\mu_\mu +\frac{1}{4}b_{\mu\nu}b^{\mu\nu}} ,
\ea
\ee
\renewcommand{\arraystretch}{1}
\hskip -.2cm which can also be checked by computing the relative string
scattering amplitudes \cite{Garousi:1996ad}.

We might have expected the appearance of a mass-term for the graviphoton 
as the presence of the brane breaks translational invariance. The gravi\-photon
indeed becomes massive and eats the $U(1)$ part of the branons, but this is not
manifest with the parametrization of the metric that we used,
$g_{MN}=\eta_{MN}+h_{MN}$. Actually, with this parametrization, the field
$h_{\mu m}$ is \emph{not} the graviphoton unless one is restricted to the
lowest order approximation. The graviphoton $V_\mu^{\ m}$ is 
defined by parametrizing the metric in the following way
\bdm
ds^2=g^{(10)}_{MN}dx^Mdx^N=g_{\mu\nu}dx^\mu dx^\nu+g_{mn}\pa{dx^m+V_\mu^{\ m}
dx^\mu}\pa{dx^n+V_\nu^{\ n}dx^\nu}\ ,
\edm
or equivalently
\be \label{menpa}
g_{MN}=\pa{\ba{cc} g_{\mu\nu} + g_{mn}V^m_\mu V^n_\nu & \ V_{\mu n} \\ 
V_{m\nu} & \ g_{mn} \ea}\ .
\ee 
Then $V_\mu^{\ m}$ can be identified with the graviphoton since the ten
dimensional coordinate transformation
\bdm
x^m\to {x^m}'=x^m+\xi^m
\edm
becomes equivalent to the gauge transformation
\bdm
V_\mu^{\ m}\to {V_\mu^{\ m}}'=V_\mu^{\ m}+\dpa_\mu\xi^m\ .
\edm
The resulting bulk kinetic terms for the dimensionally reduced theory, 
omitting the terms involving graviscalars and dilaton, is
\be \label{lbulk}
\lag_{bulk}=\frac{1}{2\kappa^2}\sqrt{|g|}\paq{\mathcal{R}^{(p+1)}-\frac{1}{4}
g_{mn}\pa{\dpa_\mu V_\nu^m-\dpa_\nu V_\mu^m}\pa{\dpa^\mu V^{\nu n}-\dpa^\nu 
V^{\mu n}}}
\ee

Expanding the Born-Infeld action (\ref{bie}) over the metric (\ref{menpa})  we
obtain, up to quadratic order in the fields
\renewcommand{\arraystretch}{1.5}
\be \label{eat}
\ba{ll} 
\lag_{2NS^2}'=-T_p\displaystyle \sqrt{|g|} &\left[ \frac{1}{2}
\pa{h^\mu_\mu+\frac{p-3}{2}\phi}+\frac{1}{2}\pa{V_\mu^m+\dpa_\mu\si^m}^2+
\frac{1}{8}\pa{h_\mu^\mu+\frac{p-3}{2}\phi}^2-\right.\\ \displaystyle
&\left.\frac{1}{4}h_{\mu\nu}h^{\mu\nu} +\frac{1}{2}\si^i\dpa_i
\pa{h^\mu_\mu+\frac{p-3}{2}\phi}
+\frac{1}{4}\pa{b_{\mu\nu}+2\pi\al F_{\mu\nu}}^2\right]
\ea
\ee
\renewcommand{\arraystretch}{1}
\hskip -.2cm where we arranged the terms involving the branons and the
graviphotons in a perfect square, showing that for each $m$ the $U(1)$ branon
is eaten by the corresponding graviphoton which becomes massive
\cite{Sundrum:1998sj}. Its mass $m_{gp}$ is given by\footnote{In our analysis
we implicitly assumed that the Kaluza-Klein scale $1/R \gg m_{gp}$
otherwise bulk terms may induce mixing and mass terms of comparable strength to
(\ref{mgp}).} 
\be \label{mgp}
m^2_{gp}=\frac{16\pi T_p}{\pa{M_{Pl}}^{p-1}}\, ,
\ee
where $M_{Pl}$ is the lower dimensional Planck mass on the $p$-brane. 
This eating mechanism is T-dual of the mechanism which makes the 
antisymmetric tensor $b_{\mu\nu}$ massive by eating the $U(1)$ world-volume
gauge field $A_\mu$ \cite{Kalb:yc}, triggered by the last term in (\ref{eat}).
In fact, a massless two-index antisymmetric tensor in $(p+1)$ dimensions has 
$\frac{1}{2}(p-1)(p-2)$ components which absorbs the $(p-1)$ components of
a gauge field through the last term of (\ref{eat}) 
to make a massive antisymmetric tensor field with $\frac{1}{2}p(p-1)$ 
components.

The terms quadratic in $h_\mu^\mu$ and $h_{\mu\nu}$ are due to the 
cosmological constant term associated to the brane tension.
The additional interaction $\si^i\dpa_i (h^\mu_\mu+(p-3)\phi/2)$ can be 
interpreted as a mixing between the longitudinal mode of the graviphoton 
(involving only the branon with the identity $\id$ Chan-Paton factor) and the
Kaluza-Klein excitations of the zero helicity part of the graviton and 
dilaton. Using canonically normalized fields, this mixing is given by
\be \label{mix1}
\lag'_{mix}=-\frac{1}{2\sqrt{16\pi}} m_{gp}M^{\frac{p-1}{2}}_{Pl} \si^i\dpa_i
\pa{h^\mu_\mu+\frac{p-3}{2}\phi}\ .
\ee
We note that there is also a similar mixing with excitations of the 
R-R sector, whose amplitude is equal in magnitude and opposite in sign to the 
previous one. The equality of the magnitude of the NS-NS and R-R mixing 
contributions is not surprising, as unitarity relate these mixing amplitudes to
the imaginary part of the one loop two branon point-function, which must vanish
in a  supersymmetric background.

We will now focus on the NS-NS sector.
The mixing (\ref{mix1}) vanishes on-shell unless $\si^m$ is massive for some
direction $\bar m$ and involves only the identity $\id$ Chan-Paton factor.
For non trivial Chan-Paton factors, we start from the trilinear coupling we
found in Sections 2, 3
\be \label{qua}
\lag=-\frac{1}{4}\si^m\si^n \dpa_m\dpa_n\pa{h^\mu_\mu+\frac{p-3}{2}\phi}-
\pa{\dpa_\mu\si^n}\si^m\dpa_mh^\mu_{\ n}
\ee
and assuming that one of the branons gets a mass $m_\si$ and a non-vanishing
VEV $v$, we substitute 
\be \label{miexp}
\si^{\bar m}=v+\rho^{\bar m}
\ee
and obtain the mixing term 
\be \label{mix2}
\lag_{mix}=-\frac{1}{2}v \rho^{\bar m}\dpa^2_{\bar m}\pa{h^\mu_\mu+
\frac{p-3}{2}\phi}+v\rho^{\bar m}\dpa_\mu\dpa_{\bar m}h^\mu_{\ \bar m}\, .
\ee

Using next the contractions
\be \label{hhff}
\rnode{g1}{h_\mu^\mu}\rnode{g2}{h^\nu_\nu} + 
\pa{\frac{p-3}{2}}^2\rnode{d1}{\phi}\rnode{d2}{\phi} &=& 
\frac{64\pi}{M_{Pl}^{p-1}}\frac{1}{k^2+m_{KK}^2}
\ncbar[nodesep=2pt,angle=-90,armA=4pt,armB=4pt]{g1}{g2} 
\ncbar[nodesep=2pt,angle=-90,armA=4pt,armB=4pt]{d1}{d2} \\
\label{pgfgf}
\rnode{r1}{h_{\bar m}^\mu}\rnode{r2}{h^\nu_{\bar n}} &=& 
\frac{16\pi}{M_{Pl}^{p-1}}
\frac{g^{\mu\nu}g_{\bar m\bar n}}{k^2+m_{KK}^2}
\ncbar[nodesep=2pt,angle=-90,armA=4pt,armB=4pt]{r1}{r2}\\ 
\rnode{a1}{h_\mu^\mu}\rnode{a2}{h^\nu_{\bar m}} &=& 0
\ncbar[nodesep=2pt,angle=-90,armA=4pt,armB=4pt]{a1}{a2} 
\ee
and assuming that $m_\si\gg 1/R$ (large extra-dimensions), so that the branons 
do not resolve the discreteness of the Kaluza-Klein spectrum, we obtain the
following correction to the branon self-energy (in the notation of
\cite{Giudice:2000av}):
\be \label{sigma}
\Sigma(p^2)=v^2\frac{16\pi V}{M^{p-1}_{Pl}}
\int \frac{d^{\delta}k}{(2\pi)^\de} 
\frac{k^4_{\bar m}+k^2k_{\bar m}^2}{p^2+k^2+i\ep}\ ,
\ee
where $\delta$ is the number of large extra dimensions and $V$ their volume.
$\Sigma$ contains the contribution from the insertion of KK modes in the branon
propagator, which reads:
\be
G_\si(p^2)=-\frac{1}{p^2+m_\si^2+\Sigma(p^2)+i\ep}\, .
\ee

The imaginary part of $\Sigma$ above is related to the decay amplitude $\Ga$ of
the  branon. Using
\bdm
\lim_{\ep\to 0}\mbox{Im}\paq{\frac{1}{x+i\ep}}=\pi \delta(x)\ ,
\edm
and the type I relation for the theory on a $p+1$ brane in ten dimensions
\cite{Sakai:1987jg}
\be \label{mplms}
M_{Pl}^{p-1}=\frac{2}{\alpha^2_{YM}}M_s^{5-p}\bar V\pa{2\pi}^{p-3}\, ,
\ee
with $\bar V$ the reduced volume defined by
\bdm
\bar V\equiv M_s^{9-p}\Pi_{i=p+1}^{9}R_i\ ,
\edm
and $M_s$ the string scale, we have
\be \label{amdis}
\Ga=\frac{1}{m_\si}\mbox{Im}\paq{\Sigma(p^2=m_\si^2)}=
\frac{4\pi\alpha^2_{YM}}{(2\pi)^{p-3}}\frac{a(\delta)}2
m_\si\pi \frac{v^2}{M_s^{5-p}}\pa{\frac{m_\si}{M_s}}^{\delta}S_{\delta-1}\, ,
\ee
where $S_{\delta-1}$ is the volume of the ($\delta-1$)-dimensional sphere of
unit radius and we defined 
\be \label{adelta}
a(\delta)\equiv \frac{3}{\delta(\delta+2)}+\frac{1}{\delta}=
\frac{\delta +5}{\delta(\delta +2)}\, .
\ee
Equation (\ref{amdis}) is the same with the expression that
appears in \cite{Giudice:2000av}, using
$M_D^{3p+\de-7}=(2\pi)^{p-3}M_s^{p+\de-1}/(4\pi\alpha_{YM}^2)$ and 
identifying $\kappa\xi$ with $\sqrt{a(\delta)/2}$. The two terms in the 
expression (\ref{adelta}) of $a(\delta)$ correspond to the contributions from
the mixing with the graviscalars (first term) and with the graviphotons (second
term).

Thus, we see that despite the absence of a $\xi\si^2\mathcal{R}$ term in the 
effective
action, a mixing can nevertheless take place with an effective $\xi$ given by
\be
\xi=\sqrt{\frac {\de+5}{6\de(\de -1)}}\, .
\ee
This mixing becomes maximal for the case of $\delta=2$ large extra dimensions,
where $\xi=\sqrt{7/12}\simeq 0.76$, leading to a possible observable invisible
width for the Higgs \cite{Giudice:2000av}. For $\delta >2$, the effective $\xi$
decreases and varies between $\xi\simeq 0.47$ for $\delta=3$ and 
$\xi\simeq 1/4$ for $\delta=6$.

\section{Higgs on branes intersection} \label{nd}

In this Section, we study the case where the Higgs lives on a branes
intersection, corresponding to an open string with mixed Neumann-Dirichlet (ND)
boundary conditions in four internal directions. We will distinguish two
subcases, depending on whether one of the two orthogonal branes extend (partly)
in the bulk of large extra dimensions.

We thus consider the coupling between two ND open string modes and a closed
string NS-NS state. We will consider first the oriented theory.
As we cannot use now the Born-Infeld action, a string calculation is the only 
way to compute this coupling. 

The kinematics of the problem is the same with the one described in
(\ref{kine}). The vertex operator for a NS open string state $\chi$, with one
end on  D5-branes and the other end on D9-branes, is (in the $(-1)$-ghost
picture, $\vf$ is the superghost field):
\be \label{v-159}
V^{(-1)}_{59}=g_ot_{aa'}\chi_\alpha e^{-\vf}\Delta S^\alpha  e^{ikX}\, ,
\ee
where the Chan-Paton factor index $a(a')$ transforms in the
(anti-)funda\-mental  of the D$5(9)$-branes gauge group. The operator $\Delta$
is the product of twist fields associated to the four internal coordinates with
mixed ND boundary conditions, $S^\alpha$ is the corresponding spin field, and
$\chi_\alpha$ selects the internal spinor helicity. This vertex operator has
the same expression as the left (supersymmetric) part of the vertex for a
massless heterotic twisted state of a $\zn_2$ orbifold \cite{Dixon:1986qv}.
For a 95 state, one has the same operator with $\chi_\alpha$ replaced by 
$\pa{\bar \chi}^\alpha\equiv \pa{\chi_\alpha}^\dagger$ and $S^\alpha$ replaced
by $S_\alpha$. The NS-NS closed string state vertex operator (in the 0-ghost
picture) is 
\be \label{2ns200}
\ba{ll}
\displaystyle V^{(0,0)}_{2NS^2}(\zeta,k)=\\ 
\displaystyle \qquad -\frac{2g_c}{\al}\zeta_{MN}\pa{i\dpa X^M +\frac{\al}{2}k
\cdot\psi\psi^M}e^{ikX} \pa{i\bar\dpa \tl X^N +\frac{\al}{2}k
\cdot\tl\psi\tl\psi^N}e^{ik\tl X}\ ,
\ea
\ee
where $X^M$ denote the bosonic coordinates and $\psi^M$ their (2d) fermionic
superpartners. 

The relevant correlators between the twist field $\Delta$ and $X$ is (for
left-movers) \cite{Hashimoto:1996he}:
\be
\frac{\langle\Delta(z_1)\Delta(z_2)X^M_L(z_3)X^N_L(z_4)\rangle}
{\langle\Delta(z_1)\Delta(z_2)\rangle}=-\frac{\al}{2}\eta^{MN}\ln\paq{
\frac{1-\sqrt\frac{z_{13}z_{24}}{z_{14}z_{23}}}
{1+\sqrt\frac{z_{13}z_{24}}{z_{14}z_{23}}}}\ ,
\ee
where $z_i$ denote the corresponding world-sheet positions.
For right-movers, the correlator is the same provided one substitutes $L,z_i$
with $R,\bar z_i$. The correlator between two $\Delta$'s is
\be
\langle \Delta(z_1)\Delta (z_2)\rangle=\frac{1}{\pa{z_1-z_2}^{1/2}}\ .
\ee

Note that the normalization coefficient $g_o$ (it is understood $g_o$ for
$p=5$) in front of the vertex operator (\ref{v-159}) is the same with the
normalization of untwisted open string states. This can
be checked  by comparing the
$\chi^2A_\mu^2$ amplitude and the exchange interaction 
$\chi A \rnode{c1}{\chi} \rnode{c2}{\chi} A\chi 
\ncbar[nodesep=2pt,angle=-90,armA=3pt,armB=3pt]{c1}{c2}$
which leads to internal propagation of a ND state.
The $\chi_\alpha$ field carries an index which labels the spinor 
representation of the internal $SO(4)$ and the GSO projection forces it to be
a Weyl spinor. Hence, it has two helicity states forming the fundamental
representation of $SU(2)$ (usually called $SU(2)_R$), rather then
the full $SO(4)$. 
This representation is pseudoreal, \emph{two-dimensional} in the complex sense
and \emph{four-dimensional} when viewed over the real numbers. In the oriented 
theory the two $\chi$'s correspond to the two independent excitations 
described by 59 and 95 states which together make up the bosonic content of an 
$\nn=1$ hypermultiplet in six dimensions.

In the unoriented theory, we expect just one complex boson (the bosonic content
of half of a hypermultiplet) as 59 and 95 modes are correlated. This is 
obtained \cite{Witten:1995gx} via a projection which involves $SU(2)_R$ as 
well as the 5-brane gauge index, being the gauge group $Sp(k)$. The projection 
is a reality condition which can be applied to the spinor $\chi$ as the 
representation $(\mathbf{2k},\mathbf{2})$ of $Sp(k)\times SU(2)_R$ is real 
(the $\mathbf{2k}$ of $Sp(k)$ being also pseudoreal).

The relevant correlators involving the spinor fields $S^\alpha$ in $4$ internal
dimensions are \cite{Kostelecky:1986xg}:
\renewcommand{\arraystretch}{1.4}
\bse
\be
&\sca{S_\alpha(z_1)S^\beta(z_2)\psi^M\psi^N(z_3)}=-\frac{1}{2}
\pa{\Ga^{MN}}_\alpha^{\ \beta}\displaystyle\frac{z_{12}^{1/2}}{z_{31}z_{32}}\\
&\ba{l}
\sca{S_\alpha(z_1)S^\beta(z_2)\psi^M(z_3)\psi^N(z_4)}=\frac{1}{2}
\pa{z_{32}z_{42}z_{31}z_{41}}^{-1/2}z_{12}^{-1/2}z_{34}^{-1}\times \\
\paq{\de^{MN}\delta_\alpha^\beta\pa{z_{32}z_{41}+z_{31}z_{42}}-
\pa{\Ga^{MN}}_\alpha^{\ \beta}z_{12}z_{34}}
\ea\\
&\ba{l}
\langle S_\alpha(z_1)S^\beta(z_2)\psi^M\psi^N(z_3)\psi^R\psi^S(z_4)\rangle=
-\displaystyle\frac{\delta^\beta_\alpha}{z_{12}}
\frac{\eta^{MR}\eta^{NS}-\eta^{MS}\eta^{NR}}{z_{34}^2}+\\
\pa{\eta^{MR}\Ga^{NS}+\eta^{NS}\Ga^{MR}-\eta^{MS}\Ga^{NR}-
\eta^{NR}\Ga^{MS}}_\alpha^{\ \beta}
\pa{2z_{34}}^{-1}\pa{z_{32}z_{42}z_{31}z_{41}}^{-1/2}\ ,
\ea
\ee
\ese
\renewcommand{\arraystretch}{1}
\hskip -.15cm where $-i \Ga_{MN}/2=-i[\Ga_M,\Ga_N]/4$ is the Lorentz generator 
in the spinor representation. This correlators can be used to compute the 
3-point amplitude, which in the $\al t\to 0$ limit becomes:
\renewcommand{\arraystretch}{1.4}
\be
\ba{l}
A_{2ND,NS^2}=2ig_c\pi\left[\pa{-k_2\cdot k_3\eta^{\mu\nu}+k_2^\mu k_3^\nu+
k_3^\mu k_2^\nu+\displaystyle \frac{2}{\pi t}
\pa{k_3^\mu k_2^\nu-k_3^\nu k_2^\mu}}\delta_\alpha^\beta\zeta_{\mu\nu}\right.\\
\left.\displaystyle +\frac{k_{1r}}4
\pa{\pa{\Ga^{rm}}_\alpha^{\ \beta}(k_2-k_3)^\nu \zeta_{m\nu}+
\pa{\Ga^{rn}}_\alpha^{\ \beta}\pa{k_2-k_3}^\mu\zeta_{\mu n}}\right]
\pa{\chi^2_\alpha}^\dagger \chi^3_\beta\, ,
\ea
\ee
\renewcommand{\arraystretch}{1}
\hskip -.2cm where $\chi^{2,3}$ is the $\chi$-field with momentum $k_{2,3}$. 
The amplitude\footnote{Note that in contrast to the $SO(3,1)$
case, for Euclidean $SO(4)$ spinors, the quantity $\chi^\dagger\xi$ is scalar
provided $\chi$ and $\xi$ have the \emph{same} chirality.} above displays
a pole term in $t$ due to the $\chi\chi \rnode{a1}{A} \rnode{a2}{A}b 
\ncbar[nodesep=2pt,angle=-90,armA=4pt,armB=4pt]{a1}{a2}$ exchange interaction 
that has to be subtracted in order to extract the contact terms. Using 
(\ref{go}) and after the usual rescaling (\ref{resca}b) one obtains the 
trilinear Lagrangian:
\renewcommand{\arraystretch}{1.7}
\be \label{lagchi}
\ba{ll}
\lag_{2ND,NS^2}=-\displaystyle
\frac{1}{2}\ &\left[-\dpa_\mu\bar\chi\dpa_\nu\chi h^{\mu\nu}+\displaystyle
\dpa\bar\chi\dpa\chi\pa{\frac{h^\mu_\mu}2+\frac{p-3}{4}\phi}+\right.\\ 
&\displaystyle\quad\frac{1}{4}\dpa_nh_{\mu m}\pa{\dpa^\mu\bar\chi \Ga^{mn}\chi-
\bar\chi \Ga^{mn}\dpa^\mu\chi}\Big]\ .
\ea
\ee
\renewcommand{\arraystretch}{1}

No contact interaction with $b_{\mu\nu}$ is found, neither a potential coupling
to the internal components of the Riemann tensor, as in the untwisted DD case
we studied in Sections 2, 3. However, besides the standard 
kinetic terms we find still a coupling of the ND open string modes to the KK
excitations of the graviphoton, arising through the spin connection in
the gravitational covariant derivative
\be
\na^{gr}_\mu\chi=\dpa_\mu\chi+\frac{1}4\om_\mu^{\ mn}\Ga_{mn}\chi\, .
\ee
Here, $\om_\mu^{\ mn}$ is the standard spin connection (with one index 
parallel and two orthogonal to the D5-brane) which is given in 
terms of the vielbein $e_\mu^a$ by
\bdm
\om_\mu^{\ mn}=\frac{1}{2}e^{\nu m}\pa{\dpa_\mu e_\nu^n-\dpa_\nu e_\mu^n}-
\frac{1}{2}e^{\nu n}\pa{\dpa_\mu e_\nu^m-\dpa_\nu e_\mu^m}
-\frac{1}{2}e^{\rho m}e^{\si n}\pa{\dpa_\rho e_{\si i}-\dpa_\si e_{\rho i}}
e^i_\mu
\edm 
whose first order expansion around flat space, $g_{MN}=\eta_{MN}+h_{MN}$, is
\be
\om_\mu^{\ mn}=h_\mu^{\ [m,n]}\ .
\ee
The connection part of the covariant derivative is completed by gauge terms
to make the full covariant derivative
\be
\na_\mu\chi=\dpa_\mu\chi+\frac{1}4\om_\mu^{\ mn}\Ga_{mn}\chi+\pa{ig_5A_\mu-
ig_9A'_\mu}\chi\ ,
\ee
where $A_\mu (A'_\mu)$ is the D5 (D9) world-volume gauge field with gauge
coupling $g_5$ ($g_9$). Finally, open string excitations $\si$ and $\chi$ have
also non-derivative ($D$-terms) interactions \cite{Polchinski}
\be \label{4cf}
\lag_D=-\frac{g^2_{YM}}{4}\left(\paq{\si^m,\si^n}-\frac i2\bar\chi\Ga^{mn}
\chi\right)^2
\ee
in the normalization of (\ref{lagchi}).

We consider now the possibility of mixing between $\chi$ and closed 
string modes. In the case where the Higgs, identified with $\chi$, lives on an
intersection of two orthogonal branes, both transverse to the submillimeter 
bulk (e.g. D3 and D7, or D5 and D5'), no mixing is generated between $\chi$ and
closed string states. On the other hand, in the case where one of the two
orthogonal branes extends in the bulk, a mixing is induced, as can be seen from
the effective Lagrangian (\ref{lagchi}), between $\chi$ and the longitudinal
component of the corresponding graviphoton in the bulk. As in the DD case, in
order to obtain a quadratic coupling between closed and open string states, one
of the $\chi$'s must acquire a non-vanishing vacuum expectation value. 

Note that a VEV of $\chi$ along a supersymmetric flat direction, i.e. when the
$D$-term (\ref{4cf}) vanishes, gives rise to the well-known Higgs branch which
provides a string  realization of a non-abelian soliton \cite{Witten:1995gx}
that we are not interested in here. We consider instead a real vacuum
expectation value for $\chi$, with non-vanishing $D$-term that breaks
supersymmetry, and we study the effective field theory obtained by expanding
around the VEV $v$, $\chi_1=v+\chi'_1$, where $\chi_1$ is one of the two
complex bosons. Dropping the prime from $\chi_1'$ and assuming the ND
conditions to be along the directions $\hat 6\ldots\hat 9$ (orthogonal to the
5-brane), we have up to quadratic order in $\chi$ and $\si$:
\be \label{mchi}
\ba{c} \displaystyle
\lag_D'=-\frac{g_{YM}^2}4\left[v^4+4v^3\mbox{Re}(\chi_1)+v^2
\pa{4(\mbox{Re}\chi_1)^2+2\chi_1^\dagger\chi_1+3\chi_2^\dagger\chi_2}+\right.\\
\displaystyle\left.+v^2\pa{[\si^{\hat 6},\si^{\hat 9}]+
[\si^{\hat 7},\si^{\hat 8}]}\right]\ ,
\ea
\ee
where fields are canonically normalized.
Note the appearance of a cosmological constant and of a tadpole for 
$\chi_1$. This is anyway only an effective approach and other potential terms 
may be generated when supersymmetry is broken.

On the other hand, $\chi$ can also obtain a mass in a supersymmetric
way, avoiding the cosmological constant and tadpole-like terms, as in
(\ref{mchi}). This is achieved by turning on a Wilson line for the gauge fields
with polarization parallel to the 5-branes, or if we T-dualize, by separating
lower and higher dimensional branes by giving an expectation value to one (or
some) of the branons orthogonal to both branes. This corresponds to moving in
the so-called Coulomb branch of the theory.

The $\chi_1$ expectation value determines the mixing terms between 
the $\chi$ field and the corresponding graviphoton 
\be
\ba{c} \displaystyle
\lag_{mix}=-\frac{1}{4}v
\pa{\dpa_{[\hat 6}h_{\hat 9]\mu}+\dpa_{[\hat 7}h_{\hat 8]\mu}}
\dpa^\mu\mbox{Im}\chi_1\, .
\ea
\ee
Using (\ref{pgfgf}), one finds for the bosonic field Im$\chi_1$,
\be 
\Sigma_\chi(p^2)=\frac{v^2}{8}\frac{8\pi V}{M^{p-1}_{Pl}}
\int \frac{d^{\delta}k}{(2\pi)^\de} \frac{k^2k_{\bar m}^2}{p^2+k^2}
\ee
and consequently, using (\ref{mplms}), one finds the following invisible width
\be
\Ga_\chi =\frac{1}{m_\chi}\mbox{Im}\paq{\Sigma(p^2=m^2_\chi)}=
\frac{4\pi\alpha_{YM}^2}{(2\pi)^{p-3}}
\frac{\pi}{32\delta}\frac{v^2}{M_s^{5-p}}m_\chi
\pa{\frac{m_\chi}{M_s}}^\delta S_{\delta -1}\ .
\ee
Hence, the resulting effective parameter $\xi$ in this case reads
\be
\xi=\frac{1}{4}\sqrt\frac{\de +2}{6\de(\de -1)}\, ,
\ee
which is significantly smaller than in the DD case, studied in Section 5.
Indeed, the highest value obtained for $\delta=2$ is $\xi\simeq 1/7$. Due to
the fact that the graviphoton and its KK tower form a quasi-continuum set of
states, this result is not altered if we consider unoriented type I models in
which  an orbifold projection takes the zero mode of the graviphoton out of 
the spectrum.

In conclusion, in this work, we investigated the possibility of mixing between
the Higgs, identified as an open string excitation, and closed string states
from the bulk (graviscalars), when the fundamental string scale is in the TeV
region. We found that such a mixing can occur, leading to a possible observable
invisible decay width of the Higgs, only when the Higgs lives on the Standard
Model world-brane and corresponds to a DD open string with both ends on 
parallel
D-branes.

As we mentioned in the introduction, although our analysis was done in the
context of supersymmetric type I theory, our results remain valid in non
supersymmetric D-brane models where supersymmetry is broken (mainly) in the
open string sector, using appropriate combinations of branes with
(anti)-orientifolds that preserve different amount of supersymmetries. The
reason is that in these cases, the effective action can be obtained by a
corresponding truncation of a supersymmetric action.

\section*{Acknowledgments}
This work was partly supported by the European Union under the RTN contracts
HPRN-CT-2000-00148, HPRN-CT-2000-00152 and the INTAS contract N 99 1 590. 
We thank F. Hassan and R. Rattazzi for discussions.
R.S. acknowledges the hospitality of the Centre de Physique Th\'eorique of the
Ecole Polytechnique, of the Oxford Theoretical Physics Department, and of the
Theory Division of CERN. During the completion of this work, R.S. was partially
supported by the COFIN grant from MURST.

\end{document}